\documentstyle[12pt,epsfig,amssymb]{article}
\def\d{{\rm d}}
\def\kp{k\cdot p}
\def\ppk{p_{+}\cdot k}

\begin{document}
\setlength{\parskip}{0.45cm}
\setlength{\baselineskip}{0.75cm}
\begin{titlepage}
\begin{flushright}
DESY-97-118 \\ DO-TH 97/14 \\ June 1997
\end{flushright}
\vspace{0.6cm}
\begin{center}
\Large
\hbox to\textwidth{\hss
{\bf The Bethe--Heitler Process in} \hss}

\vspace{0.1cm}
\hbox to\textwidth{\hss
{\bf Polarized Photon--Nucleon Interactions} \hss}

\vspace{1.2cm}
\large
T.\ Gehrmann\\
\vspace{0.5cm}
\normalsize
DESY, Theory Group,\\
\vspace{0.1cm}
D-22603 Hamburg, Germany\\
\vspace{1.2cm}
\large
M.\ Stratmann\\
\vspace{0.5cm}
\normalsize
Universit\"{a}t Dortmund, Institut f\"{u}r Physik, \\
\vspace{0.1cm}
D-44221 Dortmund, Germany \\
\vspace{1.6cm}
%
{\bf Abstract}
%
\vspace{-0.3cm}
\end{center}
We calculate the Bethe--Heitler cross section for 
the production of lepton pairs in the field of a longitudinally
polarized nucleon, taking into account the lepton masses and the target 
mass. This process is a dominant background to the detection of 
open charm from semi-leptonic decay modes, which is a 
potential probe of the polarized gluon distribution in the nucleon.  
\end{titlepage}
\noindent
The knowledge on the spin structure of the nucleon has improved 
considerably over the past few years. More precise measurements 
of the spin asymmetry $A_1(x,Q^2)\simeq g_1(x,Q^2)/F_1(x,Q^2)$ in
longitudinally polarized deep-inelastic scattering (DIS) of leptons 
off proton, deuterium and neutron targets~\cite{mallot} yielded valuable 
information on spin sum rules and on the polarized valence quark 
distributions. 
On the theoretical side, 
it has become possible to
perform a consistent analysis of polarized DIS in
next-to-leading order (NLO), since the required spin-dependent
two-loop splitting functions have been calculated recently~\cite{split}.
Nevertheless, all NLO analyses \cite{nloana1,nloana2} have demonstrated, 
that the
available data sets are still not sufficient for an accurate extraction of
the spin-dependent sea quark and gluon 
densities of the nucleon. This is true in particular
for the detailed $x$-{\em{shape}} of the spin-dependent gluon 
distribution, even though a
tendency towards a sizeable positive {\em{total}} gluon polarization
was found \cite{nloana1,nloana2}. The spin-dependent gluon distribution
enters the polarized structure 
functions at leading order (LO) only
indirectly via the $Q^2$-dependence of $g_1$ 
which could not be studied accurately up to now due to
the rather limited kinematical coverage in $(x,Q^2)$ of the
present fixed target experiments \cite{mallot}. 
Moreover, a direct extraction of the 
gluon distribution from scaling violations of the polarized structure 
functions is more involved than in the unpolarized 
case, as a complicated interplay of quark and gluon contributions to
the scaling violations~\cite{lowx} is taking place even at low values of $x$.
Clearly, the determination of the polarized gluon distribution
 is one of the most
interesting challenges for future spin physics experiments.

Recently much effort was devoted to examine the feasibility of
such measurements at future polarized $pp$  (RHIC,~\cite{rhic}) and
$ep$  colliders, one conceivable option for a future HERA upgrade
which is currently under discussion \cite{heraspin}. An 
alternative measurement could be possible  
at the recently approved COMPASS experiment~\cite{compass} at CERN
or at the proposed E156 experiment~\cite{slac} at SLAC.
The key process studied in these latter fixed target experiments  
is the production of charmed particles, as  
the cross section asymmetry for open charm
photoproduction
%
${\Delta \sigma^{\gamma N\rightarrow
c\bar{c}X}}/{\sigma^{\gamma N \rightarrow c\bar{c}X}}
$
%
provides a clear tool to access the spin-dependent gluon distribution
due to the dominance of the 
photon-gluon fusion subprocess $\gamma g \rightarrow c \bar{c}$.
Such a measurement at fixed target energies has originally been suggested
in the literature in \cite{charm1} and was further studied in \cite{charm2}.
The charm production induced by partons in the photon (the 'resolved' 
subprocess, where also the yet 
experimentally unknown polarized parton distributions of the 
{\em{photon}} enter) is moreover
shown to be negligibly small at the energies 
available at fixed target experiments
$(\sqrt{S_{\gamma N}}\lesssim 20\,\mathrm{GeV})$ for realistic
scenarios of the photonic parton densities \cite{resolved}.

The charmed events can either be detected via their hadronic 
$D$-meson decays ($D^0\rightarrow K^- \pi^+$, 
$D^{*+}\rightarrow D^0\pi_{\mathrm{soft}}^+\rightarrow 
(K^-\pi^+)\pi_{\mathrm{soft}}^+,\;\ldots$), which allow for an
efficient background rejection provided a sufficiently good
particle identification and energy resolution, 
or from the observation of decay
muons.
The hadronic $D$-meson decay channels were used in the recent
H1 and ZEUS measurements \cite{h1zeus} and will be employed also in the
upcoming COMPASS experiment \cite{compass}. The proposed SLAC 
experiment~\cite{slac} 
will use the muonic decay channels, which were first used  
in the measurements of charm 
photoproduction by EMC~\cite{emccharm}. 

Obviously, a good understanding of possible
background processes yielding 
charged lepton final states is essential in the latter case. 
The calculation of asymmetries induced by one of 
the most important background processes, the photoproduction
of charged leptons with circularly polarized photons and 
longitudinally polarized nucleons
via the Bethe-Heitler (BH) mechanism~\cite{bethe} depicted in Fig.~1
is the purpose of this paper.
There is, in principle, another source of charged leptons in photon-hadron
interactions~\cite{jaffe}: the Drell-Yan (DY) process, 
where the incoming quasi-real photon can either resolve into its
hadronic content ('resolved' process) or can act as an elementary
particle ('direct' process). 
For the energies available at the 
proposed SLAC experiment ($E_{\gamma}\lesssim 50\,\mathrm{GeV} 
\Leftrightarrow \sqrt{S_{\gamma N}}\lesssim 10\,\mathrm{GeV}$), the
contribution from the DY mechanism 
is however expected to be only marginal compared to the 
BH process as can be inferred from corresponding analyses 
in the unpolarized case \cite{jaffe}. 
Moreover,
there is also experimental indication that  lepton pairs 
in unpolarized photon-nucleon collisions are 
produced predominantly by the BH process.
The NA14 experiment, which has studied the photoproduction 
of $J/\psi$-particles at $E_{\gamma}\approx 90~\mbox{GeV}$, has found 
\cite{na14} that the lepton pair continuum below the $J/\psi$ 
resonance is well described by the Bethe--Heitler process only. 

Apart from being a background process to the detection of open 
charm, the polarized BH cross section is interesting in itself as a 
probe of the polarized structure functions $g_1$ and $g_2$.
The momenta of the two muons allow for a complete reconstruction of the 
kinematics and hence a for a measurement in any desired kinematic region
(deep inelastic, resonance, elastic). This could be in particular 
relevant for a measurement of $g_1$ in the region of low $x$ and
low $Q^2$, where the SLAC electroproduction experiments suffered 
from large pionic backgrounds~\cite{bosted}.
However, more detailed studies using the 
formulae derived below will have to 
be carried out to test the feasibility of such a measurement.  
\begin{figure}[t]
\begin{center}
\epsfig{file=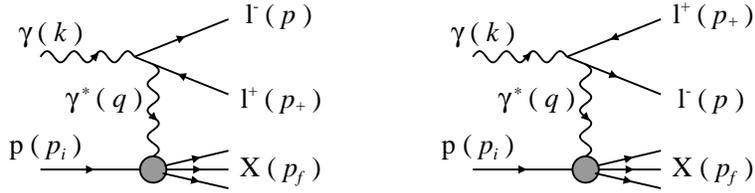,width=10cm}
\caption{Feynman diagrams for the photoproduction of 
leptons via the Bethe--Heitler process.}
\end{center}
\end{figure}

The Bethe--Heitler process in the field of a nucleon can be viewed as
the fluctuation of a real photon into an off-shell lepton--antilepton pair, 
which is put on-shell by interacting with the target nucleon, which
is not necessarily left intact (Fig.~1).
The interaction with the nucleon is described 
by the same structure functions appearing in lepton--nucleon
scattering. 
The unpolarized cross section  
$\d \bar{\sigma}$ for this process 
was originally calculated long ago by Drell and Walecka in~\cite{drell} 
and rederived 
by Kim and Tsai~\cite{tsai,tsai2}, 
whose notation we will adopt in the following. 
Moreover, polarization asymmetries in the Bethe--Heitler process 
with circularly polarized photons onto an {\em{un}}polarized target due to
electroweak interference or due to external fields have been 
derived in~\cite{singlepol}.  

We consider the production of charged lepton pairs 
(with lepton mass $m$) in the collision 
of a photon beam off a nucleon target $N$ of mass $m_i$ and spin
direction $S^{\alpha}$: 
\begin{displaymath}
\gamma(k) + N(p_i) \longrightarrow l^{+}(p_{+}) + l^{-}(p) + X (p_f)\;\;\;.
\end{displaymath}   

The hadronic final state $X$ has an invariant mass $m_f$ and the four-momentum
transfer to the target is denoted by 
$q^{\alpha} = p_f^{\alpha} - p_i^{\alpha} = k^{\alpha}
- p^{\alpha} - p_{+}^{\alpha}$ with $-q^2\equiv Q^2>0$.

The hadronic tensor is defined in terms of two spin-independent structure
functions $W_{1,2}$ and two spin-dependent ones, $G_{1,2}$, appearing in the
symmetric and anti-symmetric part of $W^{\mu\nu}$, respectively:
\begin{eqnarray}
W^{\mu\nu} & = & \frac{1}{4 \pi m_i e^2}
\sum_f \langle p_i,S | J^{*\mu}(0) |f \rangle \langle f | J^{\nu}(0) |p_i,S
\rangle \left( 2\pi \right)^4 \delta^4 \left(q+p_i-p_f\right) \nonumber \\ 
& = & -\left( g^{\mu\nu} - \frac{q^{\mu}q^{\nu}}{q^2}
\right) W_1(\nu,q^2) + \frac{1}{m_i^2} \,\left(p_{i}^{\mu} 
- \frac{p_i \cdot q}{q^2}
q^{\mu} \right) \,\left(p_i^{\nu} - \frac{p_i\cdot q}{q^2}
q^{\nu} \right) \, W_2(\nu,q^2) \nonumber \\
& & + \frac{i}{m_i^2} \epsilon^{\mu\nu\rho\sigma}
q_{\rho} \left( S_{\sigma} \left( G_1(\nu,q^2) +\frac{p_i\cdot
q}{m_i^2}\,G_2(\nu,q^2) \right)
- \frac{S\cdot q }{m_i^2}\, p_{i\sigma} G_2(\nu,q^2) \right) \;\;,
\label{eq:tensor}
\end{eqnarray}
where $\nu=p_i\cdot q/m_i$.

We define the longitudinally polarized and unpolarized 
cross sections in the usual way via 
\begin{equation}
\d \Delta \sigma \equiv \frac{1}{2}\left( \d \sigma^{\stackrel{\rightarrow}{\Leftarrow}}
- \d \sigma^{\stackrel{\rightarrow}{\Rightarrow}}\right)\;, \qquad
 \d \bar{\sigma} \equiv \frac{1}{2} \left( \d 
\sigma^{\stackrel{\rightarrow}{\Leftarrow}}
+ \d \sigma^{\stackrel{\rightarrow}{\Rightarrow}}\right)\; ,
\end{equation}
(the arrows denote the spin directions of beam and target)
such that the measurable cross section asymmetry becomes
\begin{equation}
A = \frac{\displaystyle \d \sigma^{\stackrel{\rightarrow}{\Leftarrow}}
- \d \sigma^{\stackrel{\rightarrow}{\Rightarrow}}}{\displaystyle
\d \sigma^{\stackrel{\rightarrow}{\Leftarrow}}
+ \d \sigma^{\stackrel{\rightarrow}{\Rightarrow}}} = \frac{ \d \Delta \sigma}
{\d \bar{\sigma}} \; .
\end{equation}

We present our results in the experimentally 
relevant target rest frame 
and denote the energy of the photon in this frame by $K$ and the 
energies of the lepton and antilepton by $E$ and $E_+$, respectively.
The polarized BH cross section finally reads
\begin{equation}
\d \Delta \sigma = e^6 \, \frac{m_i}{4 (k \cdot p_i)} \, \frac{\d^3 p}{E}
\, \frac{\d^3 p_+}{E_+} \, \frac{1}{2^6 \pi^5} \, \frac{1}{q^4} 
L^A_{\mu\nu}
W_A^{\mu\nu},
\label{eq:delxmaster}
\end{equation}
where the contraction of the antisymmetric parts of the leptonic and
hadronic tensors in (\ref{eq:tensor}) is given by
\begin{eqnarray}
- \frac{m_i^2}{4} L^A_{\mu\nu} W_A^{\mu\nu} & = & 
\hspace{0.9cm}
G_1(q^2,m_f^2) \left[ \frac{H_1}{(\ppk)^2} + \frac{B_1}{(\ppk)} + C_1 
+ D_1 (\ppk) + E_1 (\ppk)^2 \right] \nonumber \\
&& + \frac{q^2}{m_i} G_2(q^2,m_f^2) \left[ \frac{H_2}{(\ppk)^2} 
+ \frac{B_2}{(\ppk)} + C_2 
+ D_2 (\ppk) \right],
\label{eq:doubleres}
\end{eqnarray}
with
\begin{eqnarray}
H_1 & = & -m^2\left[ K\, q^2 + (\kp)\,\Delta  + \frac{(\kp)^2}{K} - 
\frac{(\kp)\, q^2}{2 m_i} \right] \nonumber \\
B_1 & = & \frac{(\kp)^2}{K} + \left[ \frac{1}{K} \left(q^2-2 m^2\right) - 
\frac{q^2}{2m_i} + \Delta \right] (\kp) + \frac{K\,q^4}{2(\kp)} \nonumber \\
& & + q^2 \left[
\frac{m^2}{2m_i} - (E-K) \right] - m^2 \Delta\nonumber \\
C_1 & = & \frac{1}{K} \left[ (\kp) - 2 m^2 \right] + \frac{1}{(\kp)} \left[ 
\left( q^2 - m^2 \right)\, \left(\Delta - \frac{q^2}{2 m_i} \right) 
+ E\, q^2 \right] - \frac{1}{(\kp)^2} K\, q^2\, m^2 \nonumber \\
D_1 & = & \frac{1}{K} + \frac{1}{(\kp)} \left[ \frac{1}{K} \left( q^2 - 2 m^2 
\right) + \Delta - \frac{q^2}{2 m_i} \right] + \frac{m^2}{(\kp)^2}
\left[ \frac{q^2}{2 m_i} - \Delta \right] \nonumber \\
E_1 & = & \frac{1}{K\, (\kp)} \left[ 1- \frac{m^2}{(\kp)}\right] \nonumber \\
H_2 & = & m^2 (\kp) \nonumber \\
B_2 & = & \frac{(\kp)}{K} \left[ E-K+\Delta-\frac{q^2}{2m_i}\right] + m^2 -
\frac{q^2}{2} \nonumber \\
C_2 & = & \frac{q^2}{2\, m_i\, K}-\frac{\Delta}{K} -1 + \frac{1}{(\kp)}\left[
m^2 -\frac{q^2}{2} \right] \nonumber \\
D_2 & = & \frac{1}{(\kp)} \left[ \frac{m^2}{(\kp)} -\frac{E}{K} \right]\;
\end{eqnarray} 
where we have introduced $\Delta\equiv(m_f^2-m_i^2)/(2m_i)$.

Finally, we calculate 
the cross section where only {\em{one}} of the leptons, say the $l^-(p)$, is
observed by integrating over $d^3 p_{+}$ in Eq.\ (\ref{eq:delxmaster}). 
This integration is most conveniently performed in the frame where  
$\vec{k}-\vec{p}$ is at rest and both the vectors $\vec{k}$ and $\vec{p}$ 
lie in the $xz$ plane \cite{tsai2}. 
Vectors, momenta and energies in 
this special 
frame are denoted by a subscript $s$. The angle between  $\vec{k}_s$ 
and the $z$ axis is denoted by $\Theta_k$, the angle between $\vec{p}_{+s}$ 
and the $z$ axis by $\Theta_{+}$. The projection of $\vec{p}_{+s}$ onto the 
$xy$ plane and the $x$ axis forms the angle $\phi$. 
The chosen frame has the advantage that the integration over $\phi$ 
can be straightforwardly carried out analytically, following closely the
unpolarized calculation of Kim and Tsai~\cite{tsai,tsai2}, where more details 
can be found.

By defining an auxiliary vector
\begin{displaymath}
U^{\alpha} \equiv p^{\alpha}_{+} + p_f^{\alpha} = k^{\alpha} + p_i^{\alpha}
- p^{\alpha}, 
\end{displaymath}
with 
\begin{displaymath}
U = \sqrt{U^2} = \sqrt{m^2 + m_i^2 + 2m_i(K-E) - 2 (\kp)}
\end{displaymath}
in the target rest frame, all particle energies, momenta and angles in the 
special frame can be expressed in terms of observables as follows:
\begin{eqnarray}
K_s &=& \frac{K m_i - (\kp)}{U} \nonumber \\
E_{+s} &=& \frac{U^2 + m^2 - m_f^2}{2U} \nonumber \\
p_{+s} &=& \sqrt{ E^2_{+s} - m^2} \nonumber \\
p_{is} &=& \frac{m_i}{U} \sqrt{K^2 + E^2 - m^2 + 2 \left[ (\kp) - K E\right]}
\nonumber \\ 
E_s &=& \frac{(\kp) + m_i\,E -m^2 }{U} \nonumber \\
\cos \Theta_k & = & \frac{K_s-E_s}{p_{is}} + \frac{(\kp)}{K_s\, p_{is}}
\nonumber \\ 
\cos \Theta_{+} & = & \frac{1}{p_{+s}\, p_{is}}\left[ \frac{q^2}{2}
- m^2 +  (\kp) +E_{+s} (K_s - E_s) \right]. \nonumber 
\end{eqnarray} 
The cross section (\ref{eq:delxmaster})
for detecting only the lepton $l^-(p)$ finally reads ($t=-q^2$):
\begin{eqnarray}
\frac{\d \Delta \sigma}{\d \Omega \d p } & = & 
- \frac{\alpha^3}{2 \pi m_i^2}\, \int_{m_i^2}^{(U-m)^2} \d m_f^2 \; 
\int_{t_{min}}^{t_{max}} 
\d t \;
\frac{p^2}{K\,E\,U\,q^4\,p_{is}}\nonumber \\
& & \hspace{-0.5cm}
\times \Bigg[ \hspace{0.83cm}
G_1(q^2,m_f^2) \left\{ H_1\, \frac{W}{Y^3 K_s^2} + B_1\, 
\frac{1}{Y K_s} + C_1 + D_1\, K_s W +E_1\, K_s^2 \frac{3W^2-Y^2}{2} \right\}
\nonumber\\
& & \hspace{-0.1cm}
+\frac{q^2}{m_i}\,G_2(q^2,m_f^2) \left\{ H_2\, \frac{W}{Y^3 K_s^2} + B_2\, 
\frac{1}{Y K_s} + C_2 + D_2\, K_s W \right\} \Bigg],
\label{eq:singleres}
\end{eqnarray}
where we have introduced
\begin{eqnarray}
W&=& E_{+s} - p_{+s} \cos \Theta_{+} \cos \Theta_k \nonumber\\
Y&=& \sqrt{m^2\sin^2\Theta_k + \left( p_{+s}\cos \Theta_{+} - E_{+s} \cos
\Theta_{k} \right)^2} \;\;. \nonumber
\end{eqnarray}
The integration limits $t_{min,max}$ in (\ref{eq:singleres}) are given by
\begin{displaymath}
t_{max \atop min} = - 2 m^2 + 2 (\kp) + 2 E_{+s} (K_s -E_s) \pm 2p_{+s}p_{is} .
\end{displaymath}

For completeness, it should be mentioned that the deeply inelastic limit of 
the formfactors 
$G_1$ and $G_2$ appearing in 
Eqs.\ (\ref{eq:tensor}), (\ref{eq:doubleres}), and (\ref{eq:singleres})
is related to the commonly used polarized 
structure functions $g_1$ and $g_2$ by
\begin{equation}
\frac{\nu}{m_i} G_1(\nu,q^2) = g_1(x,Q^2)\;,\;\;
\left(\frac{\nu}{m_i}\right)^2 G_2(\nu,q^2) = g_2(x,Q^2)\;\;.
\end{equation}
Finally, the elastic contribution to these form factors, relevant 
in particular for 
the calculation of the single lepton inclusive cross section  
(\ref{eq:singleres}) can be expressed as~\cite{elff}
\begin{eqnarray}
G_1^{{\rm el}}(\nu,q^2) & = & \frac{G_M(q^2)}{2\,(1+\tau)} \left[
G_E(q^2) + \tau G_M(q^2) \right]\,m_i 
\delta \left(\nu + \frac{q^2}{2m_i}\right)
\; ,\nonumber \\
G_2^{{\rm el}}(\nu,q^2) & = & \frac{G_M(q^2)}{4\,(1+\tau)}
\left[ G_E(q^2)-G_M(q^2)\right]\,
m_i \delta \left(\nu + \frac{q^2}{2m_i}\right)
\; , 
\end{eqnarray}
with $\tau=-q^2/(4m_i^2)$ and $G_{E,M}$ being the elastic nucleon form 
factors. Using the dipole parameterization~\cite{tsai2} 
\begin{displaymath}
G_E(q^2) = \frac{1}{\mu} G_M(q^2) = \left(1-q^2/(0.71\;\mbox{GeV}^2)
\right)^{-2}, 
\end{displaymath}
with $\mu$ representing the magnetic moment of the 
nucleon,  these form factors 
yield good agreement with the experimental data on 
longitudinally polarized elastic electron-proton scattering~\cite{elexp}.

The formulae (\ref{eq:delxmaster}) and (\ref{eq:singleres}) derived above 
enable, in combination with Eqs.(2.1) and (2.7) of~\cite{tsai2},  
a complete calculation of the Bethe--Heitler cross section 
with two (one) observed leptons in polarized photon--nucleon 
collisions, including all effects of lepton and target masses. They 
are in a form similar to the unpolarized cross sections~\cite{tsai,tsai2} and 
can be readily implemented into Monte Carlo simulations of 
lepton production in photon--nucleon collisions, relevant for realistic 
background estimates to the detection of open charm -- and hence to
a measurement of the polarized gluon distribution~\cite{slac}. 

Moreover, the Bethe--Heitler cross section with two detected leptons
 can in principle be used for a 
measurement of the polarized structure functions $g_1$ and $g_2$ in 
any desired kinematics. 
The feasibility of such a measurement is still to be demonstrated, and the 
formulae derived in this paper can be used to test the sensitivity of 
the polarized Bethe--Heitler process on the polarized structure functions 
for realistic experimental kinematics.

In summary, we have presented a complete calculation of the 
Bethe--Heitler photoproduction of lepton pairs in the field of 
a longitudinally polarized nucleon and given analytic expressions,
including all lepton and target mass terms, for the pair production 
and the single lepton inclusive cross sections. 

\section*{Acknowledgements}
We are grateful to P.\ Bosted for drawing our attention to the 
relevance of the polarized Bethe--Heitler process and for numerous
helpful discussions.
The work of M.S.\ has been supported in part by the
'Bundesministerium f\"{u}r Bildung, Wissenschaft, Forschung und
Technologie', Bonn.

\begin{thebibliography}{99}
%
\bibitem{mallot} A recent overview of the experimental status can be
found for example in G.~Mallot, proceedings of the ``12th International 
Symposium on High Energy Spin Physics (SPIN '96)'', Amsterdam 1996, eds.
C.W.~de Jager et al., World Scientific (Singapore, 1997), p.44.  
%
\bibitem{split} R.\ Mertig and W.L.\ van Neerven, Z. Phys. {\bf{C70}},
637 (1996);\\
W.\ Vogelsang, Phys. Rev. {\bf{D54}}, 2023 (1996); Nucl. Phys.
{\bf{B475}}, 47 (1996).
%
\bibitem{nloana1} M.\ Gl\"{u}ck, E.\ Reya, M.\ Stratmann, and
W.\ Vogelsang, Phys. Rev. {\bf{D53}}, 4775 (1996);\\
T.\ Gehrmann and W.J.\ Stirling, Phys. Rev. {\bf{D53}}, 6100 (1996);\\ 
G.~Altarelli, R.D.\ Ball, S.\ Forte, and G.\ Ridolfi,
Nucl.~Phys.~{\bf B496}, 337 (1997).
\bibitem{nloana2}
SMC Collaboration, D.~Adams et al., 
preprint CERN-PPE-97-022 (hep-ex/9702005), submitted 
to Phys. Rev. {\bf D};\\
SLAC-E154 Collaboration, K.~Abe et al., preprint SLAC-PUB-7461 
(hep-ph/9705344), to appear in Phys.~Lett.~{\bf B}.
%
\bibitem{lowx}
T.\ Gehrmann and W.J.\ Stirling, Phys. Lett. {\bf B365}, 347 (1996).
%
\bibitem{rhic} RHIC-SPIN Collaboration, M.~Beddo et al., proposal, BNL, 1992. 
%
\bibitem{heraspin} 
J.~Feltesse and A.~Sch\"afer, Proceedings of the workshop
``Future Physics at HERA'', Hamburg 1995/96, eds. G.~Ingelman, A.~De~Roeck 
and R.~Klanner, DESY (Hamburg, 1996), p.757ff. 
%
\bibitem{compass} COMPASS Collaboration, G.\ Baum et al., proposal,
CERN/SPSLC 96-14.
%
\bibitem{slac} SLAC-E156 Collaboration, 
R.G.\ Arnold et al., proposal, SLAC, 1997.  
%
\bibitem{charm1} M.\ Gl\"{u}ck and E.\ Reya, Z. Phys. {\bf{C39}}, 569
(1988).
%
\bibitem{charm2} G.\ Altarelli and W.J.\ Stirling, Particle World {\bf{1}},
40 (1989);\\
M.\ Gl\"{u}ck, E.\ Reya, and W.\ Vogelsang, Nucl. Phys. {\bf{B351}}, 579
(1991);\\
S.I.\ Alekhin, V.I.\ Borodulin, and S.F.\ Sultanov, Int. J. Mod. Phys. 
{\bf{A8}}, 1603 (1993);\\
S.\ Keller and J.F.\ Owens, Phys. Rev. {\bf{D49}}, 1199 (1994).
%
\bibitem{resolved} M.\ Stratmann and W.\ Vogelsang, Z. Phys. {\bf{C74}}, 641
(1997). 
%
\bibitem{h1zeus} H1 Collaboration, C.~Adloff et al., Z.~Phys. {\bf C72},
593 (1996);\\
ZEUS Collaboration, J.~Breitweg et al., Phys. Lett. {\bf B401}, 192 (1997).
\bibitem{emccharm}
EMC Collaboration, J.J.~Aubert et al., Nucl. Phys. {\bf B213}, 31 (1983).
%
\bibitem{bethe}
H.A.~Bethe and W.~Heitler, Proc.~R.~Soc. {\bf A146}, 83 (1934). 
%
\bibitem{jaffe} R.L.\ Jaffe, Phys. Rev. {\bf{D4}}, 1507 (1971);\\
A.\ Vourdas, J. Phys. {\bf{G6}}, 789 (1980);\\
J.\ Busenitz and J.D.\ Sullivan, Phys. Rev. 
{\bf{D24}}, 1794 (1981).
%
\bibitem{na14} NA14 Collaboration, R.~Barate et al., Z. Phys. {\bf C33}, 505
(1987). 
%
\bibitem{bosted} P.~Bosted, private communication.
%
\bibitem{drell} S.D.\ Drell and J.D.\ Walecka, Ann. Phys. (NY)
{\bf{28}}, 18 (1964).
%
\bibitem{tsai} K.\ Kim and Y.-S.\ Tsai, Phys. Lett. {\bf{40B}}, 665 (1972).
\bibitem{tsai2}
Y.-S.~Tsai, Rev. Mod. Phys. {\bf{46}}, 815 (1974), Erratum {\bf 49}, 
421 (1977).
%
\bibitem{singlepol} V.M.\ Kuznetsov and A.P.\ Potylitsyn, Sov. J. Nucl.
Phys. {\bf{27}}, 79 (1978);\\
H.\ Konashi, K.\ Ushio, and Y.\ Yokoo, Prog. 
Theor. Phys. {\bf{62}}, 1062 (1979).
%
\bibitem{elff}
A.I.~Akhiezer et al., Sov. Phys. JETP {\bf 6}, 588 (1958);\\
N.~Dombey, Rev. Mod. Phys. {\bf 41}, 236 (1969);\\
T.V.~Kuchto and N.M.~Shumeiko, Nucl.~Phys. {\bf B219}, 412 (1983).
%
\bibitem{elexp}
M.J.~Alguard et al., Phys. Rev. Lett. {\bf 37}, 1258 (1976).
%
\end{thebibliography}
\end{document}